\documentstyle[aps,epsfig,twocolumn]{revtex}

\begin{document}

\title{Comment on   ``A Kac-potential treatment of nonintegrable 
interactions" by Vollmayr-Lee and Luijten}
\author{Constantino Tsallis \\
Centro Brasileiro de Pesquisas Fisicas,
Rua Xavier Sigaud 150 \\ 
22290-180 Rio de Janeiro-RJ, Brazil  (tsallis@cbpf.br)}

\maketitle

\begin{abstract}
We comment the recent manuscript by Vollmayr-Lee and Luijten 
[cond-mat/0009031] focusing on classical systems with nonintegrable 
interactions. The authors claim that they have proved that 
Boltzmann-Gibbs statistics suffices for describing the system in 
thermal equilibrium. We show that this statement is a misleading 
oversimplification since it only applies for the $\lim_{N \rightarrow 
\infty} \lim_{t \rightarrow \infty}$ ordering, but certainly not for 
the $\lim_{t \rightarrow \infty} \lim_{N \rightarrow \infty}$ one. 
The latter can even be the {\it unique} physically meaningful 
situation for thermodynamically large systems.

\end{abstract}

\bigskip

In a recent paper \cite{VLL}, Vollmayr-Lee and Luijten (VLL) present 
a Kac-potential approach of nonintegrable interactions. They consider 
a $d$-dimensional classical fluid with two-body interactions 
exhibiting a hard core as well as an attractive potential 
proportional to $r^{-\tau}$ with $0 \le \tau/d < 1$ (logarithmic 
dependance for $\tau/d = 1$)\cite{notation}. In their approach, they 
also include a Kac-like long-distance cutoff $R$ such that no 
interactions exist for $r>R$, and then discuss the $R \rightarrow 
\infty$ limit. They show that the exact solution within 
Boltzmann-Gibbs statistical mechanics is possible and that -- no 
surprise (see VLL Ref. [12] and references therein) -- it exhibits a 
mean field criticality. Moreover, the authors argue that very similar 
considerations hold for lattice gases, $O(n)$ and Potts models. 

Almost as an illustration of the fact that {\it ``It is the strange 
privilege of statistical mechanics to stimulate and nourish 
passionate discusions related to its foundations [...]"} 
\cite{nicolis}, VLL state {\it ``Our findings imply that, contrary to 
some claims, Boltzmann-Gibbs statistics is sufficient for a standard 
description of this class of nonintegrable interactions."}, and also 
that {\it ``we show that nonintegrable interactions do not require 
the application of generalized $q$-statistics."}.  VLL also inform us 
that {\it ``the main motivation for this} [their] {\it work stems 
from the considerable attention systems with nonintegrable 
interactions have received in the context of ``nonextensive 
thermodynamics.""}.  

It is the purpose of the present Comment to argue that VLL's (first 
two) above statements severely misguide the reader. Indeed, the 
interesting VLL discussion, along traditional lines, of their 
specific Kac-like model only exhibits that Boltzmann-Gibbs 
statistical mechanics is -- as more than one century of brilliant 
successes guarantees! -- {\it necessary} for obtaining thermal 
equilibrium properties {\it without} needing to do time averages; by 
no means it proves that it is {\it sufficient}, as we shall soon 
clarify. Neither it proves that wider approaches (such as, for 
instance, nonextensive statistical mechanics, VLL Refs. [6,31] and 
present Ref. \cite{tsallis}, or a similar formalism) are not required 
or convenient. The crucial point concerns {\it time}, a word that 
nowhere appears in the VLL paper. This is, in fact, rather surprising 
since the key role of $t$ has been strongly emphasized in several 
occasions, for instance in VLL Ref. [31] (e.g, Fig. 4 of of VLL Ref. 
[31] illustrates the expectation). For integrable (or ``short-range", 
as frequently referred to in the literature focusing the present 
context \cite{longrange}) two-body interactions in a $N$-body 
classical Hamiltonian system, i.e., for $\tau/d > 1$, we expect that 
the $t \rightarrow \infty$ and $N \rightarrow \infty$ limits are 
commutable  in what concerns the equilibrium distribution $p(E)$, $E$ 
being the total energy level associated with the macroscopic system. 
More precisely, we expect naturally that 
\begin{eqnarray}
p(E) &\equiv& \lim_{t \rightarrow \infty} \lim_{N \rightarrow \infty} 
p(E;N;t) 
= \lim_{N \rightarrow \infty} \lim_{t \rightarrow \infty} p(E;N;t)  
\nonumber \\
&\propto& \exp[-E/kT]  \;\;\;(\tau/d >1)
\end{eqnarray} 
if the system is in thermal contact with a thermostat at temperature 
$T$. In contrast, the system is expected to behave in a more complex 
manner for nonintegrable (or ``long-range") interactions, i.e., for 
$0 \le \tau/d \le 1$. In this case, no generic reasons seem to exist 
for the $t \rightarrow \infty$ and $N \rightarrow \infty$ limits to 
be commutable, and consistently we expect not necessarily equal 
results. The simplest of these results (which is in fact the one to 
be associated with the VLL paper, although therein these two relevant 
limits and their ordering are not mentioned) is, as we shall soon 
further comment,
\begin{equation}
\lim_{N \rightarrow \infty} \lim_{t \rightarrow \infty}    p(E;N;t) 
\propto \exp[-(E/{\tilde N})/(kT/{\tilde N})] 
\end{equation} 
where we have introduced ${\tilde N} \equiv 
[N^{1-\tau/d}-\tau/d]/[1-\tau/d]$ in order to stress the facts that 
{\it generically} 

(i) $E$ is {\it not} extensive, i.e., is {\it not} proportional to 
$N$ (but is $E \propto N{\tilde N}$ instead; more precisely, $E$ is 
extensive if $\tau/d >1$, see \cite{fisher} and VLL Refs. [4,5], and 
it is nonextensive if $0 \le \tau/d \le 1$), 

and 

(ii) $T$ needs to be rescaled (a feature which is frequently absorbed 
in the literature by artificially size-rescaling the coupling 
constants of the Hamiltonian), in order to guarantee nontrivial {\it 
finite} equations of states. Of course, for $\tau = 0$, we have 
${\tilde N}=N$, which recovers the traditional Mean Field scaling. 

But, {\it depending on the initial conditions}, which determine the 
time evolution of the system if it is assumed isolated, quite {\it 
different} results can be obtained for the ordering $\lim_{t 
\rightarrow \infty} \lim_{N \rightarrow \infty}    p(E;N;t)$. This 
fact has been profusely detected and stressed in the related 
literature (see, for instance, VLL Ref. [31], present Refs. 
\cite{tsallis,posch,antonitorcini,ruffoetal,latorarapisarda,koyamakonishi} 
and references therein). 

Let us discuss this point further. The impressive success of 
Boltzmann-Gibbs statistical mechanics relies on the fact that, for 
ubiquitous Hamiltonian systems, time averages (whose calculation is 
typically untractable) can be replaced by appropriate ensemble 
averages (by far less harder to calculate). More specifically, let us 
consider an isolated $N$-body system (microcanonical ensemble) for 
which we fix the total energy $E$ (for some systems, other constants 
of motion need to be fixed as well). If we assume an initial 
condition for the system and let it evolve (following Newtonian 
mechanics if the system is a classical one), after sufficiently long 
time $t$ (mathematically speaking, in the limit $t \rightarrow 
\infty$; practically speaking, for times well above the inverse 
maximum Lyapunov exponent, whenever it is positive) the distribution 
of energies is given by an expression which is the microcanonical 
equilibrium distribution. From this distribution, we can in principle 
calculate the marginal distribution associated to any {\it one} among 
the $N$ particles. 
This equilibrium distribution typically is, as first perceived by 
Gibbs, a {\it power-law} (see, for instance, \cite{posch,plastinos}). 
For fixed value of $E$ properly scaled with $N$ and $N \rightarrow 
\infty$, this distribution approaches the celebrated Boltzmann, {\it 
exponential} factor, which corresponds in fact to the canonical 
ensemble, i.e., when the system is in thermal contact with a 
thermostat (which fixes $T$, of course). A beautiful illustration is 
explicitely worked out in \cite{posch} for the marginal distribution 
corresponding to the momentum of one particle: in the $N \rightarrow 
\infty$ limit, the celebrated Maxwellian velocity distribution is 
neatly recovered. By the way, in 1993 Plastino and Plastino remarked 
\cite{plastinos} that the microcanonical equilibrium power-law 
distribution just mentioned precisely is the one which emerges from 
nonextensive statistical mechanics where $1-q$ plays essentially the 
role of $1/N$.

What we have just described corresponds clearly to the \\ 
$\lim_{N \rightarrow \infty} \lim_{t \rightarrow \infty}    p(E;N;t)$ 
ordering. What relevant modifications are expected to happen in the  
$\lim_{t \rightarrow \infty} \lim_{N \rightarrow \infty}    p(E;N;t)$ 
ordering? Typically none {\it if} no long-range interactions are 
involved. But, as mentioned above, the situation is expected to be 
much more subtle in the presence of long-range interactions. To be 
more precise, consistently with the available numerical results, we 
generically expect (as conjectured in Fig 4 of VLL reference [31], 
and exhibited in 
\cite{posch,antonitorcini,ruffoetal,latorarapisarda}, among others) 
that, after some transient towards equilibration starting from 
certain classes of initial conditions (such as waterbag or double 
waterbag in velocities, for instance), the system achieves a {\it 
quasi-stationary} or {\it meta-equilibrium} state whose duration {\it 
diverges} with $N$. After this state, for finite $N$, the system 
typically relaxes to the Boltzmann-Gibbs equilibrium (i.e., $q=1$). 
Let us stress at this point that, if $N \simeq 10^{23}$, during times 
which could be longer than the age of the universe, it is the 
anomalous, non Boltzmann-Gibbsian, meta-equilibrium state which will 
be observed and {\it not} the standard one (focused in the VLL 
paper). In other words, between the standard and such nonstandard 
macroscopic states, something analogous to an activation barrier 
exists, whose height diverges with $N$. A typical (very illustrative 
but by no means unique) such system is the classical $d$-dimensional 
model of localized planar rotators with two-body (attractive) 
interactions which decay as $r^{-\tau}$, and whose coupling constant 
does {\it not} (unphysically) depend on $N$. The following anomalies 
have been (either analytically or numerically) observed in the 
microcanonical molecular dynamics approach:

(i) {\it Above} a rescaled critical energy  $u_c \equiv U_c/(N 
{\tilde N})$ where $U$ denotes the fixed total energy ($u_c=0.75$ for 
$\tau=0$), the system exhibits a (conveniently scaled) Lyapunov 
spectrum whose largest Lyapunov exponent approaches, for increasingly 
large $N$, a finite positive value for $\tau/d>1$, but vanishes 
instead for $0 \le \tau/d \le 1$ (it vanishes, in fact, as 
$N^{-\kappa}$, where $\kappa(\tau/d)$ seems to be an universal 
function, independent of both $d$ and $u$, which decreases from 1/3 
to zero when $\tau/d$ increases from zero to unity, see VLL reference 
[35] and present reference \cite{giansanti});   

(ii) {\it Below} $u_c$, long standing metaequilibrium states are 
observed everytime the simulation is started from single (or even 
double) waterbag distributions in velocities. Specifically, the time 
evolution of the (properly size-scaled) average kinetic energy (also 
averaged over many statistically equivalent sets of initial 
conditions), exhibits \cite{latorarapisarda} plateaux corresponding 
to temperatures appreciably different from the one associated with 
the canonical Boltzmann-Gibbs distribution. The duration of these 
plateaux seems to {\it diverge} with $N$, while their associated 
height remains {\it finite}. This result carries a most important 
corollary: it strongly suggests that the zeroth principle of 
thermodynamics can be valid {\it out} from the usual ($q=1$) 
statistical mechanics \cite{latraptsal};

(iii) Consistently with point (ii), the one-particle distribution of 
momenta is {\it not} the Maxwellian one during the entire extent of 
the plateaux just mentioned. It does not change for all times 
included in the plateaux, it is more peaked than the Maxwellian one 
for low momenta (consistently with a nonextensive statistical 
mechanics canonical distribution for $q>1$), and exhibits a cut-off 
for high momenta, as generically expected for any microcanonical 
distribution (indeed, since the total energy is finite, even if all 
the energy was momentarily concentrated in the kinetic energy of a 
single particle, the corresponding momentum could not be infinite);

(iv) Consistently with points (ii) and (iii), a L\'evy-type anomalous 
superdiffusion is observed during the plateaux, which makes a 
crossover to normal diffusion when the plateau ends, and the usual 
Boltzmann-Gibbs equilibrium is attained.

It is very clear that no reason at all exists for expecting the above 
list to be exhaustive in describing the anomalies associated with the 
meta-stable states. It has been given just as an illustration of the 
important features that the VLL paper has overlooked. It is in this 
sense that we have argued in the beginning of this Comment that the 
VLL paper does not prove, contrary to what is therein stated, that 
Boltzmann-Gibbs is {\it sufficient}, it only provides one more 
illustration that it is {\it necessary}, as probably all mechanical 
statistical physicists of the world are convinced. Do we mean by this 
that we have herein proved that, for classical conservative many-body 
long-range interacting Hamiltonians, nonextensive statistical 
mechanics is also {\it necessary}? Certainly not! What we believe 
that we have shown is that another, nonstandard formalism is also 
needed: at the present stage, for the particular system we are 
focusing on, nonextensive statistical mechanics is but a candidate 
(perhaps the correct one), on which many scientists are currently 
working. In a kind of desperate last effort, one could easily imagine 
a devil's advocate arguing that Boltzmann-Gibbs statistical mechanics 
is the uniquely correct one for describing the {\it ultimate} 
equilibrium macroscopic state. On this we have no objection at all, 
it even strongly seems that this is a stricly correct statement. 
However, for long-range interacting macroscopic systems initially 
placed in a nonstandard bassin of attraction (in the space of the 
distributions) of the initial conditions, this state might be 
achieved ``after the end of the universe". This feature would make 
its physical utility a very controversial matter, of a byzantine 
style which is not much of our taste. Along these lines, the only 
thing that really matters is whether, yes or no, a neat connection 
can be established between the power-law distributions which 
extremize the nonextensive entropic form $S_q \equiv [1-\sum_i 
p_i^q]/[q-1]$ \cite{tsallis} and the long standing metaequilibrium 
states existing {\it before} the ultimate, Boltzmann-Gibbs 
equilibrium state is attained. To achieve this goal, a tractable way 
is to perform microcanonical molecular dynamics simulations in 
long-range-interacting systems which contain $N>>1$ particles, and 
then focusing on subsystems of them with $M>>1$ particles such that 
$N>>M>>1$. In the $(M,N/M) \rightarrow (\infty,\infty)$ limit the 
comparison is expected to make sense. Work along this possibility is 
in progress.

It is certainly relevant to remind at this point that the celebrated 
Boltzmann-Gibbs exponential distribution for Hamiltonian systems has 
been founded in the literature in at least four manners, namely the 
variational entropic principle \cite{gibbs}, the steepest descent 
method \cite{darwinfowler}, the laws of large numbers \cite{khinchin} 
and the microcanonical counting \cite{balian}. All four have been 
generalized now (respectively \cite{tsallis}, \cite{aberajagopal1}, 
\cite{aberajagopal2} and \cite{aberajagopal3}), and systematically 
lead to the power-law distributions emerging within nonextensive 
statistical mechanics.

Although perhaps not totally transparent, the VLL paper and 
consistently the present Comment focus {\it essentially} on classical 
Hamiltonian many-body systems. However, the concepts involved in 
nonextensive statistical mechanics have been successfully applied to 
many other systems. These include fully developed turbulence 
\cite{beck,arimitsu} (Beck has recently developed a quite impressive 
theory with not a single free parameter, which compares remakably 
well with the available experimental data), granular matter 
\cite{granular}, electron-positron annihilation and other high energy 
systems \cite{highenergy}, L\'evy and correlated types of anomalous 
diffusions \cite{anomalousdiff}, low-dimensional nonlinear dynamical 
systems \cite{maps}, self-organized criticality \cite{SOC}, 
reassociation in folded proteins \cite{bemski}, quantum entanglement 
\cite{entanglement}, to quote but a few. For these systems, it is 
allowed to think that the question which remains to be fully 
clarified is {\it not} whether the formalism works succesfully, but 
{\it why} it does so. Given the wide diversity of the systems under 
focus, it is not yet totally clear what are the basic ingredients of 
the game. It is however already acquired that some type of 
(multi)fractality or hierarchical structure is apparently always 
present (the physical mechanisms capable of driving such fractality 
appear to be long-range interactions, strongly nonmarkovian 
processes, fractal boundary conditions, quantum nonlocality, 
mesoscopic dissipation, and others). This fact tends to generate a 
slow, power-law, mixing in phase and analogous spaces, as opposed to 
the usual, exponential mixing. The situation might well be, in some 
cases, more subtle than just this (for instance, in the system of 
long-range interacting rotators addressed above, the Lyapunov 
exponents in the $N \rightarrow \infty$ limit vanish for $u>u_c$ but 
are finite for $u<u_c$; see \cite{remark}), but it is presently 
unavoidable to think that microscopic mixing issues lay at the heart 
of nonextensive statistical mechanics and thermodynamics. This 
possibility goes in fact very well along ideas of Krylov 
\cite{krylov}, Balescu \cite{balescu}, Dorfman \cite{dorfman} and 
others. Further studies are certainly needed and welcome.

It is with pleasure that I acknowledge useful remarks by D.H. 
Zanette, S. Abe and A.K. Rajagopal.


\begin{thebibliography}{10}

\bibitem{VLL}B.P. Vollmayr-Lee and E. Luijten, cond-mat/0009031.

\bibitem{notation}In VLL Refs. [7-12,29,31,32,35] and elsewhere, 
$\tau \equiv d+\sigma$ is noted $\alpha$ (both cases $0 \le \alpha/d 
\le1$ and $\alpha/d >1$ are considered in those references; VLL 
focuse on $0 \le \tau/d \le 1$). In \cite{posch} $\tau$ is noted 
$-\nu$ (in \cite{posch}, only $\nu >0$ is considered). What VLL refer 
to as ``nonintegrable" (``integrable") interactions is referred to as 
"long-range" ("short-range") interactions in VLL Refs. [7-12 
,29,31,32,35] and elsewhere, and corresponds  to $0 \le \tau/d \le 1$ 
($\tau/d>1$).

\bibitem{nicolis}G. Nicolis and D. Daems, Chaos {\bf 8}, 311 (1998).

\bibitem{tsallis}C. Tsallis, J. Stat. Phys. {\bf 52}, 479 (1988); 
E.M.F. Curado and C. Tsallis, J. Phys. A {\bf 24}, L69 (1991) 
[Corrigenda: {\bf 24}, 3187 (1991) and {\bf 25}, 1019 (1992); C. 
Tsallis, R.S. Mendes and A.R. Plastino, Physica A {\bf 261}, 534 
(1998); the bibliography of the subject is regularly updated at 
http://tsallis.cat.cbpf.br/biblio.htm;  for recent reviews see C. 
Tsallis, in {\it Nonextensive Statistical 
Mechanics and Thermodynamics}, eds. S.R.A. Salinas and C. Tsallis, 
Braz. J. Phys. {\bf 29}, 1 (1999) [accessible at 
http://sbf.if.usp.br/WWW$_{-}$pages/Journals/BJP/Vol29/
Num1/index.htm], C. Tsallis, {\it Entropic nonextensivity: A possible 
measure of complexity}, to appear in the Proc. of the "International 
Workshop on Classical and Quantum Complexity and Nonextensive 
Thermodynamics" (Denton, Texas, 3-6 April 2000), eds. P. Grigolini, 
C. Tsallis and B.J. West, Chaos , Solitons and Fractals (2001) [Santa 
Fe Institute Working Paper 00-08-043 (2000); cond-mat/0010150], and 
C. Tsallis,  in {\it Nonextensive Statistical Mechanics and its 
Applications}, eds. S. Abe and Y. Okamoto, Series {\it Lecture Notes 
in Physics} (Springer-Verlag, Berlin, 2000), in press.

\bibitem{longrange} The word {\it integrable} is used, in the present 
context, when $\int dr r^{d-1} r^{-\tau}$ is finite. It is at this 
point worthy reminding that the expressions ``short" and ``long" 
range interactions are used in different senses in different chemical 
and physical areas. VLL address this fact stating that {\it ``a 
pervasive notational problem in the nonextensive thermodynamics 
literature is the use of ``long-range interactions" to mean 
``nonintegrable interactions""}; also, VLL use expressions such as 
{\it ``considerably more important"} class of integrable interactions 
and {\it ``true"} long-range interactions to communicate to the 
readers their preferences. They mention the VLL Refs. [1,27,28], to 
which one could add, along the same vein, M.E. Fisher and V.A. 
Privman, Comm. Math. Phys. {\bf 103}, 527 (1986), among others. In 
the present context, we prefer to use instead the notation commonly 
adopted in the nonextensive thermodynamics literature, which is 
consistent with L. Tisza, Annals Phys. {\bf 13}, 1 (1961) [or in 
{\it Generalized thermodynamics}, (MIT Press, Cambridge, 1966), p. 
123]. [In his words: {\it ``The situation is different for the 
additivity 
postulate P a2, the validity of which cannot be inferred from general 
principles. We have to require that the interaction energy between 
thermodynamic systems be negligible. This assumption is closely 
related 
to the homogeneity postulate P d1. From the molecular point of view, 
additivity and homogeneity can be expected to be reasonable 
approximations for systems containing many particles, provided that 
the 
intramolecular forces have a short range character."}] and with  P.T. 
Landsberg, {\it Thermodynamics and 
Statistical Mechanics}, (Oxford University Press, Oxford, 1978; also 
Dover, 1990), page 102 [In his words: {\it ``The presence of 
long-range 
forces causes important amendments to thermodynamics, some of which 
are 
not fully investigated as yet."}]. Ludwig Boltzmann himself must have 
had perceptive intuitions along related lines. Indeed, in the first 
page of the second part of his {\it Vorlesungen uber Gastheorie}, he 
qualifies the concept of ideal gas by writing: {\it " When the 
distance at which two gas molecules interact with each other 
noticeably is vanishingly small relative to the average distance 
between a molecule and its nearest neighbor -- or, as one can also 
say, when the space occupied by the molecules (or their spheres of 
action) is negligible compared to the space filled by the gas -- 
..."}.

\bibitem{fisher}M.E. Fisher, Arch. Rat. Mech. Anal. {\bf 17}, 377 
(1964); J. Chem. Phys. {\bf 42}, 3852 (1965); J. Math. Phys. {\bf 6}, 
1643 (1965).

\bibitem{posch}Lj. Milanovic, H.A. Posch and W. Thirring, Phys. Rev. 
E {\bf 57}, 2763 (1998).

\bibitem{antonitorcini}M. Antoni and A. Torcini, Phys. Rev. E {\bf 
57}, R6233 (1998).

\bibitem{ruffoetal}V. Latora, A. Rapisarda and S. Ruffo, Phys. Rev. 
Lett. {\bf 80}, 692 (1998); Physica D {\bf 131}, 38 (1999); Phys. 
Rev. Lett. {\bf 83}, 2104 (1999); Physica A {\bf 280}, 81 (2000); M. 
Antoni, S. Ruffo and A. Torcini, Proc. of the Workshop "The Chaotic 
Universe" (Rome-Pescara, February 1999) (World Scientific), in press 
[cond-mat/9908336]. 

\bibitem{latorarapisarda}V. Latora and A. Rapisarda, {\it Dynamical 
quasi-stationary states in a system with long-range forces},  to 
appear in the Proc. of the "International Workshop on Classical and 
Quantum Complexity and Nonextensive Thermodynamics" (Denton, Texas, 
3-6 April 2000), eds. P. Grigolini, C. Tsallis and B.J. West, Chaos , 
Solitons and Fractals  (2001) [cond-mat/0006112].

\bibitem{koyamakonishi}H. Koyama and T. Konishi, cond-mat/0008208 and 
cond-mat/0008507.

\bibitem{plastinos}A.R. Plastino and A. Plastino, Phys. Lett. A {\bf 
193}, 140 (1994).

\bibitem{giansanti}A. Campa, A. Giansanti, D. Moroni and C. Tsallis, 
{\it Long-range interacting classical systems: universality in mixing 
weakening}, cond-mat/0007104.

\bibitem{latraptsal}V. Latora, A. Rapisarda and C. Tsallis, in 
progress.

\bibitem{gibbs}J.W. Gibbs, {\it Elementary Principles in Statistical 
Mechanics} (Yale University Press, New Haven, 1902).

\bibitem{darwinfowler}C. G. Darwin and R. H. Fowler, Phil. Mag. and 
J. Sci. {\bf 44}, 450 (1922); R. H. Fowler, Phil. Mag. and J. Sci. 
{\bf 45}, 497 (1923).

\bibitem{khinchin}A.I. Khinchin, {\it Mathematical Foundations of 
Statistical Mechanics} (Dover, New York, 1949). 

\bibitem{balian}R. Balian and N.L. Balazs, Ann. Phys. (NY) {\bf 179}, 
97 (1987).

\bibitem{aberajagopal1}S. Abe and A.K Rajagopal, {\it Microcanonical 
foundation for systems with power-law distributions}, J. Phys. A 
(2000), in press [cond-mat/0002159].

\bibitem{aberajagopal2}S. Abe and A.K Rajagopal, {\it Justification 
of power-law canonical distribution based on generalized central 
limit theorem}, Europhys. Lett. (2000), in press [cond-mat/0003380].

\bibitem{aberajagopal3}S. Abe and A.K Rajagopal, Phys. Lett. A {\bf 
272}, 341 (2000).
 
\bibitem{beck}C. Beck, Physica A {\bf 277}, 115 (2000); C. Beck, {\it 
Non-extensive statistical mechanics approach to fully developed 
hydrodynamic turbulence},  to appear in the Proc. of the 
"International Workshop on Classical and Quantum Complexity and 
Nonextensive Thermodynamics" (Denton, Texas, 3-6 April 2000), eds. P. 
Grigolini, C. Tsallis and B.J. West, Chaos , Solitons and Fractals  
(2001) [cond-mat/0005408].

\bibitem{arimitsu}T. Arimitsu and N. Arimitsu,  Phys. Rev. E {\bf 
61}, 3237 (2000); J. Phys. A {\bf 33}, L235 (2000); {\it Tsallis 
statistics and turbulence}, to appear in the Proc. of the 
"International Workshop on Classical and Quantum Complexity and 
Nonextensive Thermodynamics" (Denton, Texas, 3-6 April 2000), eds. P. 
Grigolini, C. Tsallis and B.J. West, Chaos , Solitons and Fractals  
(2001).  

\bibitem{granular}Y.-H. Taguchi and H. Takayasu,  Europhys. Lett. 
{\bf 30}, 499 (1995); see also A. Kudrolli and J. Henry, Phys. Rev. E 
{\bf 62}, R1489 (2000).

\bibitem{highenergy}I. Bediaga, E.M.F. Curado and J. Miranda,  
Physica A {\bf 286}, 156 (2000); C. Beck, Physica A {\bf 286}, 164 
(2000); D.B. Walton and J. Rafelski, Phys. Rev. Lett. {\bf 84}, 31 
(2000); G. Wilk and Z. Wlodarczyk, Phys. Rev. Lett. {\bf 84}, 2770 
(2000); G. Wilk and Z. Wlodarczyk, {\it The imprints of nonextensive 
statistical mechanics in high energy collisions}, to appear in the 
Proc. of the "International Workshop on Classical and Quantum 
Complexity and Nonextensive Thermodynamics" (Denton, Texas, 3-6 April 
2000), eds. P. Grigolini, C. Tsallis and B.J. West, Chaos , Solitons 
and Fractals (2001) [hep-ph/0004250]; F.S. Navarra, O.V. Utyuzh, G. 
Wilk and Z. Wlodarczyk, {\it Violation of the Feynman scaling law as 
a manifestation of nonextensivity}, to appear in N. Cimento (2000) 
[hep-ph/0009165];  D.B. Ion and M.L.D. Ion, Phys. Rev. Lett. {\bf 
81}, 5714 (1998); M.L.D. Ion and D.B. Ion,  Phys. Rev. Lett. {\bf 
83}, 463 (1999); D.B. Ion and M.L.D. Ion, Phys. Rev. E {\bf 60}, 5261 
(1999); D.B. Ion and M.L.D. Ion, {\it Optimality, entropy and 
complexity for nonextensive quantum scattering},  to appear in the 
Proc. of the "International Workshop on Classical and Quantum 
Complexity and Nonextensive Thermodynamics" (Denton, Texas, 3-6 April 
2000), eds. P. Grigolini, C. Tsallis and B.J. West, Chaos , Solitons 
and Fractals (2001); M.L.D. Ion and D.B. Ion, {\it Strong evidences 
for correlated nonextensive quantum statistics in hadronic 
scatterings}, Phys. Lett. B {\bf 482}, 57 (2000);  G. Kaniadakis, A. 
Lavagno and P. Quarati, Phys. Lett. B {\bf 369}, 308 (1996); P. 
Quarati, A. Carbone, G. Gervino, G. Kaniadakis, A. Lavagno and E. 
Miraldi, Nucl. Phys. A {\bf 621}, 345c (1997); G. Kaniadakis, A. 
Lavagno and P. Quarati, Astrophysics and space science {\bf 258}, 145 
(1998); A. Lavagno, Proc. of Baryons 98 (Bonn, 22-26 September 1998), 
eds. D.W. Menze and B. Metsch (World Scientific, Singapore, 1999), 
page 709; M. Coraddu, G. Kaniadakis, A. Lavagno, M. Lissia, G. 
Mezzorani and P. Quarti, in {\it Nonextensive Statistical Mechanics 
and Thermodynamics}, eds. S.R.A. Salinas and C. Tsallis, Braz. J. 
Phys. {\bf 29}, 153 (1999); W.M. Alberico, A. Lavagno and P. Quarati, 
Eur. Phys. J C {\bf 12}, 499 (1999); A. Lavagno and P. Quarati, Nucl. 
Phys. B, Proc. Suppl. {\bf 87}, 209 (2000); A. Lavagno and P. 
Quarati, {\it Classical and quantum non-extensive statistics effects 
in nuclear many-body problems}, to appear in the Proc. of the 
"International Workshop on Classical and Quantum Complexity and 
Nonextensive Thermodynamics" (Denton, Texas, 3-6 April 2000), eds. P. 
Grigolini, C. Tsallis and B.J. West, Chaos, Solitons and Fractals  
(2001).  

\bibitem{anomalousdiff}D.H. Zanette and P.A. Alemany, Phys. Rev. 
Lett. {\bf 75}, 366 (1995); C. Tsallis, S.V.F Levy, A.M.C. de Souza 
and R. Maynard, Phys. Rev. Lett. {\bf 75}, 3589 (1995) [Erratum: {\bf 
77}, 5442 (1996)]; M. Buiatti, P. Grigolini and A. Montagnini, Phys. 
Rev. Lett. {\bf 82}, 3383 (1999); D. Prato and C. Tsallis,  Phys. 
Rev. E {\bf 60}, 2398 (1999); C. Budde, D. Prato and M. Re, preprint 
(2000) [cond-mat/0007038]; A. Robledo, Phys. Rev. Lett. {\bf 83}, 
2289  (1999); A. Robledo,  J. Stat. Phys. {\it 100}, 475 (2000); 
A.R.Plastino and A.Plastino, Physica A {\bf 222}, 347 (1995); C. 
Tsallis and D.J. Bukman, Phys. Rev. E {\bf 54}, R2197 (1996).

\bibitem{maps}C. Tsallis, A.R. Plastino and W.-M. Zheng, Chaos, 
Solitons and Fractals {\bf 8}, 885 (1997); U.M.S. Costa, M.L. Lyra, 
A.R. Plastino and C. Tsallis, Phys. Rev. E {\bf 56}, 245 (1997); M.L. 
Lyra and C. Tsallis, Phys. Rev. Lett. {\bf 80}, 53 (1998); M.L. Lyra, 
Ann. Rev. Comp. Phys. , ed. D. Stauffer (World Scientific, Singapore, 
1998), page 31; U. Tirnakli, C. Tsallis and M.L. Lyra, Eur. Phys. J. 
B  {\bf 10}, 309 (1999); U. Tirnakli, {\it Asymmetric unimodal maps: 
Some results from $q$-generalized bit cumulants}, Phys. Rev. E  (1 
Dec 2000), in press [cond-mat/9911420]; M. Buiatti, P. Grigolini and 
L. Palatella, Physica A {\bf 268}, 214 (1999); C.R. da Silva, H.R. da 
Cruz and M.L. Lyra, {\it Low-dimensional non-linear dynamical systems 
and generlized entropy}, in {\it Nonextensive Statistical Mechanics 
and Thermodynamics}, eds. S.R.A. Salinas and C. Tsallis, Braz. J. 
Phys. {\bf 29}, 144 (1999); R.S. Johal and R. Rai, Physica A {\bf 
282}, 525 (2000); V. Latora, M. Baranger, A. Rapisarda and C. 
Tsallis,  Phys. Lett. A {\bf 273}, 97 (2000); U. Tirnakli, G.F.J. 
Ananos and C. Tsallis, {\it Generalization of the Kolmogorov-Sinai 
entropy: Logistic- and periodic-like dissipative maps at the chaos 
threshold}, preprint (2000) [cond-mat/0005210]; J. Yang and P. 
Grigolini, Phys. Lett. A {\bf 263}, 323 (1999); S. Montangero, L. 
Fronzoni and P. Grigolini, {\it The non-extensive version of the 
Kolmogorov-Sinai entropy at work}, cond-mat/9911412;  F.A.B.F. de 
Moura, U. Tirnakli and M.L. Lyra, {\it Convergence to the critical 
attractor of dissipative maps: Log-periodic oscillations, fractality 
and nonextensivity}, Phys. Rev. E {\bf 62} (1 November 2000), in 
press [cond-mat/0008130]; M. Baranger, V. Latora and A. Rapisarda, 
{\it Time evolution of thermodynamic entropy for conservative and 
dissipative maps},  to appear in the Proc. of the "International 
Workshop on Classical and Quantum Complexity and Nonextensive 
Thermodynamics" (Denton, Texas, 3-6 April 2000), eds. P. Grigolini, 
C. Tsallis and B.J. West, Chaos , Solitons and Fractals (2001) 
[cond-mat/0007302].

\bibitem{SOC}F.A. Tamarit, S.A. Cannas and C. Tsallis, Eur. Phys. J. 
B {\bf 1}, 545 (1998); A.R.R. Papa and C. Tsallis, Phys. Rev. E {\bf 
57}, 3923 (1998); P.M. Gleiser, F.A. Tamarit and S.A. Cannas, Physica 
A {\bf 275}, 272 (2000).

\bibitem{bemski}C. Tsallis, G. Bemski and R.S. Mendes, Phys. Lett. A 
{\bf 257}, 93 (1999).

\bibitem{entanglement}S. Abe and A.K. Rajagopal, Physica A {\bf 289}, 
157 (2001) [quant-ph/0001085]; C. Tsallis, S. Lloyd and M. Baranger, 
{\it Generalization of the Peres criterion for local realism through 
nonextensive entropy}, quant-ph/0007112.

\bibitem{remark}To illustrate the kind of subtleties that might be 
involved, let us consider a simple example. The solution of the 
linear ordinary differential equation ${\dot \xi} = \lambda_1 \xi$ 
(with $\xi(0)=1$) is $\xi = e^{\lambda_1 t}$, where $\xi$ represents 
the sensitivity to the initial conditions and $\lambda_1$ is assumed 
to be the Lyapunov exponent of some simple, one-dimensional, 
nonlinear dynamical system. If $\lambda_1 =0$, the appropriate 
equation might be the nonlinear one ${\dot \xi} = \lambda_q\; 
\xi^q\;\;(q \in {\cal R})$, whose solution is now a power-law, namely 
$\xi=[1+(1-q)\lambda_q t]^{[1/(1-q)]}$. This solution recovers the 
previous one as the $q=1$ instance. Let us now consider the more 
general case where {\it both} linear and nonlinear terms are present. 
The corresponding differential equation can be written as ${\dot \xi} 
= \lambda_1\xi + (\lambda_q-\lambda_1) \xi^q$. The solution is now 
given by $\xi=[1-\frac{\lambda_q}{\lambda_1}+ 
\frac{\lambda_q}{\lambda_1}e^{(1-q)\lambda_1 t}]^{1/(1-q)}$. The pure 
linear case is recovered in the $\lambda_q=\lambda_1\;\;(\forall q)$ 
case as well as in the $q=1$ cases. The pure nonlinear case is 
recovered in the $\lambda_1=0$ particular case. An interesting 
situation appears when $q<1$ and $\lambda_q >>\lambda_1>0$. In this 
case, for $t<<1/[(1-q)\lambda_1]$ the solution practically coincides 
with the power-law one, and for $t>>1/[(1-q)\lambda_1$ the solution 
asymptotically reproduces the exponential behavior. Generally 
speaking, it is possible to imagine that some specific effects could 
depend basically only on  the {\it nonlinear} term ($\propto \xi^q$) 
and practically not on the linear term ($\propto \xi$). If so, the 
relevant part of $\xi(t)$ would be basically related to the power-law 
behavior, {\it whether $\lambda_1$ is zero or not}. Considerations of 
this sort might have relevance for deeply understanding what happens 
dynamically in macroscopic systems like the long-range interacting 
rotators one discussed in this Comment, where a value $u_c$ exists 
such as at {\it both} sides of which unusual features are observed: 
vanishing Lyapunov exponents for $u>u_c$, and the existence of long 
standing metastable states coexisting with nonzero Lyapunov exponents 
for $u<u_c$. A situation analogous to this one does occur in standard 
critical phenomena. Indeed, a crucially important quantity (directly 
related to the correlation length, the temperature dependance of 
which defines in turn the critical exponent $\nu$) is the two-body 
space correlation, {\it not} of the values of the local order 
parameter at different sites, but of the {\it fluctuations}, with 
regard to the average order parameter, of the values of the local 
order parameter at different sites . The distinction is inexistent 
for the disordered phase (typically at $T>T_c$), where the average 
order parameter vanishes, but it is most important for the ordered 
phase (typically at $T<T_c$), where the average order parameter is 
nonzero.  

\bibitem{krylov}N. Krylov, Nature {\bf 153}, 709 (1944) 
[In his words: {\it `` In the present investigation, the notion of 
ergodicity 
is ignored. I reject the ergodical hypothesis completely: it is both 
insufficient and unnecessary for statistics. I use, as 
starting point, 
the notion of motions of the mixing type, 
\newpage
and show that the essential 
mechanical condition for the applicability of statistics consists in 
the requirement that in the phase space of the system all the regions 
with a sufficiently large size should vary in the course of time in 
such a way that while their volume remains constant -- according to 
Liouville's theorem -- their parts should be distributed over the 
whole 
phase space (more exactly over the layer, corresponding to given 
values 
of the single-valued integrals of the motion) with a steadily 
increasing degree of uniformity. (...) The main condition of mixing, 
which ensures the fulfillment of this condition, is a sufficiently 
rapid 
divergence of the geodetic lines of this Riemann space (that is, of 
the 
paths of the system in the $n$-dimensional configuration space), 
namely, an exponential divergence (cf. Nopf$^1$). "}]. For full 
details 
on this pioneering approach see N.S. Krylov, {\it Works on the 
Foundations of Statistical Physics}, translated by A.B. Migdal, Ya. 
G. 
Sinai and Yu. L. Zeeman, Princeton Series in Physics (Princeton 
University Press, Princeton, 1979).

\bibitem{balescu}R. Balescu, {\it Equilibrium and
Non-equilibrium Statistical Mechanics} (John Wiley, New York,
1975), page 727 [In his words: {\it ``It therefore appears from the
present discussion that the {\it mixing property of a
mechanical system is much more important for the understanding
of statistical mechanics than the mere ergodicity.} (...) A
detailed rigorous study of the way in which the concepts of
mixing and the concept of large numbers of degrees of freedom
influence the macroscopic laws of motion is still lacking.}].

\bibitem{dorfman}J.R. Dorfman, {\it An Introduction to Chaos in 
Nonequilibrium Statistical Mechanics}, Cambridge Lecture Notes in 
Physics {\bf 14}, eds. P. Goddard and J. Yeomans (Cambridge 
University 
Press, Cambridge, 1999), footnote in page 9 [In his words: {\it `` It 
is 
worth mentioning that there are examples of mixing systems with no 
non-zero Liapunov exponents. The concepts of ergodicity, mixing, and 
chaos can be quite subtle"}].

\end{thebibliography}
\end{document}